\documentclass[intlimits,twoside,a4paper]{article}
\usepackage{graphicx}
\usepackage[T2A]{fontenc}
\usepackage[cp1251]{inputenc}
\usepackage[ukrainian,english]{babel}
\usepackage{amsmath,amssymb}
\usepackage{cmpj}

\issue{2011}{14}{2}{23001}

\doinumber{10.5488/CMP.14.23001}

\title[Merton-Garman model]
{The path integral representation kernel of evolution operator in
Merton-Garman model}

\author[L.F.~Blazhyevskyi, V.S.~Yanishevsky]{L.F.~Blazhyevskyi\,\refaddr{label1},
        V.S.~Yanishevsky\,\refaddr{label2}}
\addresses{
 \addr{label1} Department for Theoretical Physics, Ivan Franko National University of Lviv,
12~Drahomanov~Str.,\\  79005~Lviv, Ukraine
 \addr{label2} Department for Economical Cybernetics and Innovations,
Drohobych Ivan Franko State Pedagogical University, 46~Lesja
Ukrajinka~Str., 82100~Drohobych, Ukraine}

\authorcopyright{L.F.~Blazhyevskyi, V.S.~Yanishevsky, 2011}
\date{Received September 7, 2010, in final form February 23, 2011}

\begin{document}
\maketitle
\begin{abstract}
In the framework of path integral the evolution operator kernel
for the Merton-Garman Hamiltonian is constructed. Based on this
kernel option formula is obtained, which generalizes the
well-known Black-Scholes result. Possible approximation numerical
schemes for path integral calculations are proposed.
\pacs 03.65.-w., 89.65.Gh, 89.65.s, 02.50.r, 02.50.Cw
\end{abstract}

\section*{Introduction}

The path integral method was proposed by R.~Feynman as a new
approach~\cite{Feyn_Hib} to the solution of quantum mechanical
problems. Nowadays it became one of the most powerful methods in
theoretical physics.  Many applications
(see~\cite{Klein,Schul,Roep,Wieg}) of this method are devoted to
diverse important problems. In articles~\cite{Baaq97,Baaq}
describing physical analogies, the path integral method was
applied to a series of financial tasks. One of these tasks is the
problem of evaluating the option price, initiated in the seminal
work~\cite{Black} which is known as the Black-Scholes model (see
appendix~\ref{Bl_Sch}). Some generalization of this model is the
Merton-Garman model, where there was suggested
\cite{Baaq97,Baaq,Hull_03} a suitable dynamics equation of the
option price. Just like the Schrodinger equation, the
Merton-Garman equation is of evolution type. Hence, the path
integral method is well fit for presenting the corresponding
solution in a closed form and reduces the problem to quadratures.
In articles~\cite{Baaq97,Baaq} the path integral representation
was constructed for the operator evolution kernel (propagator) of
a Merton-Garman Hamiltonian.

Meanwhile, while constructing the evolution operator kernel
in~\cite{Baaq97,Baaq}, there were encountered a few shortcomings.
In particular,  the operator structure of the Merton-Garman
Hamiltonian was not taken into account which holds the terms with
mutually non-commuting summands by analogy with momentum and
coordinate operators of quantum mechanics. In the case under
regard, the evolution operator kernel for the Merton-Garman
equation is not determined within  the Feynman path integral (on
the phase space) from the classical Hamiltonian function entailing
from the Hamiltonian operator by replacing the momentum and
coordinate operators by their classical expressions
$\hat{p}_y\rightarrow p_y\,$ ($\hat{p}_y=-{\rm
i}\,{\partial}/{\partial y}$). Correspondingly, the same holds for
the path integral in the configuration space, and thus the
classical Lagrange function does not make it possible to correctly
determine the evolution operator kernel. The situation mentioned
above is well known~\cite{Klein,Schul,Grosch,Proh,Blaz} in quantum
mechanics and concerns the operator ordering problem. It is
known~\cite{Proh} that in the case of terms with non-commuting
factors within the Hamiltonian the corresponding action in the
path integral still contains additional components. Subject to the
Trotter formula approach (which was used in~\cite{Baaq,Baaq97}),
the above mentioned components were chosen from the correct form
of the resulting Hamiltonian evolution equation based on the
obtained evolution operator kernel. As it follows from~\cite{Proh}
(see also~\cite{Blaz}), since these additional components were not
taken into account the evolution operator kernel does not
correspond to the initial one but rather to some other operator
symmetrized over the momentum operators with the Hamiltonian
operator. Thus, the evolution operator kernel constructed
in~\cite{Baaq,Baaq97}, does not correspond to the initial
Merton-Garman Hamiltonian but to some other Hamiltonian. The
determination of these additional components within the Trotter
approach is a rather cumbersome procedure~\cite{Proh}. In our
work, we use an alternative approach for construction of the path
integral evolution operator kernel representation based on the
Gaussian path integral within the $T$-product
approach~\cite{Blaz,Nazaj} (see appendix~\ref{Tprod}). It is well
known that this approach is effective for Hamiltonians with
quadratic momentum dependence, that is the case for the
Merton-Garman Hamiltonian. We constructed the corresponding path
integral evolution kernel representation for the Merton-Garman
equation having interpreted the integration variable as the
velocity~\cite{Klein,Blaz}. Moreover, we determined the additional
components within the natural Gauss integration procedure assuming
the proper symmetrization. Based on this evolution operator kernel
we presented an analog of the Black-Scholes formula for a
European call option price.

\section{The basic Merton-Garman model setting}
Within the Black-Scholes model~\cite{Baaq, Black}  two main
financial instruments are considered: bank account and stocks. A
bank account change $\Pi(t)>0$~-- on risk-free instruments and
governed by the equation:
\begin{align}\label{accon}
\rd\Pi(t)=r\Pi(t)\rd t,
\end{align}
where the value of $r\geqslant0$~-- interest rate. A share price
change $S(t)\geqslant0$~--  can be determined from the stochastic
equation:
\begin{align}\label{asset}
\rd S(t)=\phi\,S(t)\rd t +\sigma\,S(t)\,R(t)\rd t,
\end{align}
$\phi\geqslant0$~-- is the expected return on the security $S(t)$
and $\sigma\geqslant0$~-- stock price volatility (the standard
deviation). The quantities $\phi$, $\sigma$ are taken to be
constant. The second summand~\eqref{asset} contains a Gaussian
random function $R(t)$ (the white noise), whose first moments are
equal:
\begin{align}\label{gaus:R}
\langle R(t)\rangle=0,\qquad \langle
R(t)R(t')\rangle=\delta(t-t').
\end{align}
The brackets above $\langle\ldots\rangle$ denote the Gauss average
of the random variable $R(t)$. Stochastic equation~\eqref{asset}
is called a geometric (as well as economical) Brown
motion~\cite{Schyr}. As concerns the classical Langevin equation
for the Brown particle velocity~\cite{Bales,Lind}, a random
component of equation~\eqref{asset} is proportional to $S(t)$. The
quantity $\rd W(t)=R(t)\rd t$ in equation~\eqref{asset} is also
referred to as the standard Brown motion (or the Wiener process)
which describes the independent increases of quantity $R(t)$ with
characteristics:
\begin{align*}
\langle W(t)\rangle=0,\qquad \langle W(t)W(t')\rangle=\min(t,t').
\end{align*}
\par
The basic assumptions of the classical option pricing theory are
that the option price $C(t,S(t))$ at time $t$ is a continuous
function of time and its underlying asset's price $S(t)$. To
realize this dependence, it is necessary to impose some
relationships. The Black and Scholes idea, concerning the
determination of a rational option price~\cite{Black}, was based
on a call option payment (premium) $C(t)$ at a moment of time $t$
($0\leqslant t\leqslant T$), considered by an option seller as a
minimal capital (investment portfolio), whose investments will
permit a contract to get fulfilled. Making use of this aspect,
Black and Scholes have determined~\cite{Black,Baaq} the following
investment portfolio:
\begin{align}\label{porto}
C(t,S(t))=\Pi(t)+\frac{\partial C(t,S(t))}{\partial S(t)}S(t).
\end{align}
The first summand in~\eqref{porto} means the premium part which is
placed in the bank account, the second part is assigned for buying
shares at market prices $S(t)$. Within the economical sense,
namely coefficient ${\partial C(t,S(t))}/{\partial S(t)}>0$
determines the amount of shares bought at the price $S(t)$. For
option price dynamics equation to be obtained we need to find the
change in time of both parts of equality~\eqref{porto}, having
taken into account equations~\eqref{accon},~\eqref{asset}. In
particular, in the left hand side of~\eqref{porto} one obtains
(for brevity, the $t$-argument is omitted):
\begin{align}\label{porto:1}
\frac{\rd C}{\rd t}=\frac{\partial C}{\partial t}+\frac{\partial
C}{\partial S}\frac{\rd S}{\rd t}+\frac1{2}\frac{\partial^2
C}{\partial S^2}\frac{(\rd S)^2}{\rd t}\,.
\end{align}
The last summand in~\eqref{porto:1} needs to
take~\cite{Schyr,Baaq} the relationship $(\rd S)^2=\sigma^2S^2\rd
t$ into account. Taking now the derivative from the right hand
side of~\eqref{porto} and equating it to~\eqref{porto:1}, upon
some simplifications, one obtains~\cite{Black, Schyr, Baaq} the
Black-Scholes equation:
\begin{align}\label{Blac:1}
\frac{\partial C}{\partial t}+rS\frac{\partial C}{\partial
S}+\frac1{2}\sigma^2S^2\frac{\partial^2 C}{\partial S^2}=rC.
\end{align}
Making change of variables $S=\re^x\,(x\in\mathbb{R}$) we obtain
equation \eqref{eq:BS}. Its solution (appendix \ref{Bl_Sch})
permits to obtain the Black-Scholes formula for the rational
option price~\eqref{price:BS}.

\par As it was  noticed above, the volatility $\sigma^2$ within the Black-Scholes model
is a constant value. Thus,  some generalizations of the
Black-Scholes model were proposed in which the volatility changes
with time. In particular, in a rather general form, the volatility
dynamics equation was written \cite{Hull_03, Baaq} as follows:
\begin{align}\label{V}
\rd V(t)=\left[\lambda+\mu V(t)+\xi V^{\alpha}(t) R_2(t)\right]\rd
t,
\end{align}
where $V(t)=\sigma^2(t)$, $R_2(t)$~-- Gauss white noise. There are
also conditions imposed on the constants $\lambda,\,\mu\,,\xi$
which provide the positivity $V(t)>0$. Then, the dynamics share
price can be written as follows:
\begin{align}\label{dyn:SV}
\rd S(t)=\phi S(t)\rd t+\sqrt{V(t)}S(t)R_1(t)\rd t.
\end{align}
The correlation Gauss white noise $R_1(t)$, $R_2(t)$ moments are
determined by relationships~\eqref{gaus:R}, as well as by:
\begin{align}\label{gaus:R12}
\langle R_1(t)R_2(t')\rangle=\rho\,\delta(t-t')
\end{align}
with the correlation parameter $-1\leqslant\rho\leqslant1$.

While deriving the dynamics option price equation the
Black-Scholes portfolio property~\eqref{porto} was taken into
account. It is easy to see that the relationship structure
of~\eqref{porto} gives rise to the summand contraction of
$\backsim{\rd S(t)}/{\rd t}$, and the equation~\eqref{Blac:1} does
not contain the terms proportional to the random component
$\backsim R(t)$. Meanwhile, the price $S(t)$ is determined at time
$t$ (similar to the Brown particle velocity, whose velocity is
known at initial moment $t$, but its change is a random variable).
Making use of this property while forming the investment
portfolio,  the following equation was obtained (its more detailed
analysis can be found in~\cite{Baaq, Baaq97}) for the option price
dynamics:
\begin{align}\label{Hm:MG0}
\frac{\partial C}{\partial t}+rS\frac{\partial C}{\partial S}+(\lambda+\mu
V)\frac{\partial C}{\partial V}+\frac1{2}VS^2\frac{\partial^2 C}{\partial S^2}+
\rho\xi V^{1/2+\alpha}S\frac{\partial^2 C}{\partial S\partial
V}+\xi^2V^{2\alpha}\frac{\partial^2 C}{\partial V^2}=rC.
\end{align}
Introducing new variables $S=\re^{x},\,V=\re^{y}$ ($x,\,y
\in\mathbb{R}$) equation \eqref{Hm:MG0} can be written down as
follows:
\begin{align}\label{eq:MG}
 \frac{\partial C(t)}{\partial t}=H_{\rm MG}C(t),
\end{align}
where we denoted the quantity:
\begin{align}\label{Hm:MG}
H_{\rm MG}=&-\frac{\re^y}2\frac{\partial^2}{\partial
x^2}+\left(\frac{\re^y}2-r\right)\frac{\partial}{\partial
x}+r-\left(\lambda
\re^{-y}+\mu-\frac{\xi^2}2\re^{2y(\alpha-1)}\right)\frac{\partial}{\partial y}\,-\notag\\
&-\rho\xi \re^{y(\alpha-1/2)}\frac{\partial^2}{\partial x\partial
y}-\frac{\xi^2}2 \re^{2y(\alpha-1)}\frac{\partial^2}{\partial
y^2}\,,
\end{align}
which means exactly the Merton-Garman Hamiltonian. It is easy to
see that the first three summands in~\eqref{Hm:MG} correspond to
the Black-Scholes Hamiltonian term, while the next ones are owing
to the stochastic dynamics of the volatility.

\section{The path integral representation of the Merton-Garman equation}\label{path:CM}

The continuous solutions to equation~\eqref{sol:BS2}, which is
understood in the standard generalized sense, can be represented
in the form:
\begin{align}\label{sol:MG}
C(t)=\re^{-\tau H_{\rm MG}}C(T),
\end{align}
where $C(T)$ is the value of the option price at time $t$, which
is determined by function~\eqref{prifun}. The evolution operator
kernel in~\eqref{sol:MG}, similar to~\eqref{ker:BS_2}, can be
determined by the relationship:
\begin{align}\label{Kern0:MG0}
g(\tau,x-x_0\,,y,y_0)=\re^{-\tau H_{\rm
MG}}\delta(x-x_0)\delta(y-y_0).
\end{align}

Since the Hamiltonian coefficients in~\eqref{Hm:MG} are
independent of the variable $x$, based on the experience at the
construction of the Black-Scholes Hamiltonian and its evolution
operator kernel, one can make the Fourier transformation with
respect to the $x$-variable and as a result, one obtains the
Hamiltonian~\eqref{Hm:MG} in the form:
\begin{align}\label{HmF:MG}
\tilde{H}_{\rm MG}(k)=h_0(k,y)+h_1(k,y)\frac{\partial}{\partial
y}-\frac{\xi^2}2\left(\re^{y(\alpha-1)}\frac{\partial}{\partial
y}\right)^2,
\end{align}
where we have denoted:
\begin{eqnarray}\label{par_h}
h_0(k,y)=\frac{\re^y}{2}k^2+\ri\left(\frac{\re^y}{2}-r\right)k+r,
\end{eqnarray}
\begin{eqnarray}\label{par_hhhh}
h_1(k,y)=-\ri\rho\xi \re^{y(\alpha-1/2)}k-\left(\lambda
\re^{-y}+\mu-\frac{\xi^2\alpha}{2} \re^{2y(\alpha-1)}\right).
\end{eqnarray}
Respectively, the Fourier image of the kernel~\eqref{Kern0:MG0} is
determined as:
\begin{align}\label{Kern0:MGk}
\tilde{g}(\tau,k,y,y_0)=\re^{-\tau\tilde{H}_{\rm
MG}(k)}\delta(y-y_0).
\end{align}

\subsection{The case $\alpha =1$}\label{alpha1}

Consider the simplest case $\alpha=1$ where the coefficient at the
operator ${\partial^2}/{\partial y^2}$ does not depend on the
variable $y$. Having chosen the complete square form in the
Hamiltonian~\eqref{HmF:MG}, one obtains:
\begin{align}\label{HmF2:MG}
\tilde{H}_{\rm MG}(k)=-\frac1{2}\left(\xi\frac{\partial}{\partial
y}-\frac{h(k,y)}{\xi}\right)^2+h_0(k,y)+\frac1{2}\frac{h^2(k,y)}{\xi^2}-\frac1{2}h'(k,y).
\end{align}
Here the quantity $h(k,y)$ equals the value
$h_1(k,y)$~\eqref{par_h} for $\alpha=1$:
\begin{align*}
h(k,y)&=-\ri\rho\xi \re^{y/2}k-\left(\lambda
\re^{-y}+\mu-{\xi^2}/{2}\right)
\end{align*}
and the summand $h'(k,y)$ means the derivative of the $h(k,y)$ with respect to
the variable $y$:
\begin{align*}
h'(k,y)=\lambda \re^{-y}-\ri k\frac1{2}\rho\xi \re^{y/2}.
\end{align*}
Namely, a similar summand, as we will show below, was not taken
into account in articles~\cite{Baaq, Baaq97}. In our case, this
appears to have been caused by the corresponding symmetrization
performed in~\eqref{HmF2:MG}. One can also observe that this
summand is important for the Hamiltonian \eqref{HmF2:MG} (in the
coordinate representation $k\rightarrow -\ri{\partial}/{\partial
x}$) whose form would take an incomplete form:
\begin{align}\label{HmBag:MG}
H^*_{\rm MG}=&-\frac{\re^y}2\frac{\partial^2}{\partial
x^2}+\left(\frac{\re^y}2-r\right)\frac{\partial}{\partial
x}+r-\left(\mu-\frac{\xi^2}2\right)\frac{\partial}{\partial
y}-\frac{\lambda}{2}\left(
\re^{-y}\frac{\partial}{\partial y}+\frac{\partial}{\partial y}\re^{-y}\right)\,-\notag\\
&-\frac{\rho\xi}{2}\frac{\partial}{\partial
x}\left(\re^{y/2}\frac{\partial}{\partial
y}+\frac{\partial}{\partial
y}\re^{y/2}\right)-\frac{\xi^2}2\frac{\partial^2}{\partial y^2}\,.
\end{align}
It is seen that that summands with non-commuting factors
in~\eqref{HmBag:MG} are written in a symmetrized form. Namely, as
it was already mentioned above, the path integral representation
of the evolution operator kernel, constructed in \cite{Baaq,
Baaq97}, corresponds to~\eqref{HmBag:MG} in contrast to the
initial Hamiltonian~\eqref{Hm:MG} (for $\alpha=1$).

The evolution operator in~\eqref{Kern0:MGk} can be transformed by
using the expression~\eqref{gaus:1}:
\begin{align}\label{EvOp:MG}
&\re^{-\tau \tilde{H}_{\rm
MG}(k)}=\,{\mathcal{N}}(\tau)\!\!\int\mathcal{D}v
\exp\left(-\frac1{2}\int_0^\tau v^2(t)\rd
t\right)\times\notag\\&\times T\exp\left(\int_0^\tau\left[
-h_0(k,{y})-\frac1{2\xi^2}h^2(k,y)+\frac1{2}
h'(k,{y})+\frac{1}{\xi}v(t)h(k,y)\right]\rd t-\int_0^\tau \xi
v(t)\rd t\frac{\partial}{\partial y}\right).
\end{align}
The latter summand in the $T$-exponent contains a shift-operator
which can be disentangled with respect to~\eqref{Texp}, since the
$T$-exponent becomes a usual one:
\begin{align*}
T\exp\left(-\int_0^\tau \xi v(t)\rd t\frac{\partial}{\partial
y}\right)&=\exp\left(-\int_0^\tau \xi v(t)\rd t\frac{\partial}{\partial y}\right),\\
\left[T\exp\left(-\int_0^\tau \xi v(t)\rd
t\frac{\partial}{\partial
y}\right)\right]^{-1}&=\exp\left(\int_0^\tau \xi v(t)\rd
t\frac{\partial}{\partial y}\right).
\end{align*}
Respectively, for the operator kernel~\eqref{Kern0:MGk}, upon
simple transformations which are omitted, one can obtain:
\begin{align}\label{Kern:MG}
\tilde{g}(\tau,k,y,y_0)=\,&\bar{\mathcal{N}}(\tau)\!\!\int\mathcal{D}v
\exp\left(-\frac1{2\xi^2}\int_0^\tau [v(t)-h(k,\tilde{y})]^2\rd
t-\int_0^\tau h_0(k,\tilde{y})\rd
t\right)\times\notag\\&\times\exp\left({\frac1{2}\int_0^\tau
h'(k,\tilde{y})\rd t}\right) \delta\left(y-y_0-\int_0^\tau v(t)\rd
t\right),
\end{align}
where we denoted $\tilde{y}=y-\int_t^{\tau}v(t')\rd t'$ and the
constant $\bar{\mathcal{N}}(\tau)$ is determined from the
condition:
\begin{align*}
\bar{\mathcal{N}}(\tau)\!\!\int\mathcal{D}v
\exp\left(-\frac1{2\xi^2}\int_0^\tau v^2(t)\rd t\right)=1.
\end{align*}
The path integral~\eqref{Kern:MG} is defined in the space of
velocities \cite{Klein, Blaz}. The quantity
\begin{align*}
L(k)=\frac1{2\xi^2}[v-h(k,y)]^2+h_0(k,y)
\end{align*}
means a Lagrangian function (for imaginary time) and the summand
$-h'(k,{y})/2$~--  the additional term.
Relationship~\eqref{Kern:MG} determines the Fourier image with
respect to $x$-variable. For the Fourier inverse transform one
needs to separate in~\eqref{Kern:MG} the $k$-dependence:
\begin{align}\label{Kern1:MG}
\tilde{g}(\tau,k,y,y_0)=\re^{-\tau
r}\bar{\mathcal{N}}(\tau)\!\!\int\mathcal{D}v
\re^{S_0}\exp\left(-\frac1{2}\tau\tilde{\sigma}^2k^2-\ri\tau
k\left[{\tilde{\sigma}^2}/2-\tilde{r}\right]\right)\delta\left(y-y_0-\int_0^\tau
v(t)\rd t\right),
\end{align}
where we denoted:
\begin{align}\label{Par1}
S_0&=-\frac1{2\xi^2}\int_0^{\tau}\left[ v(t)+\lambda\,{
\re^{-\tilde{y}}}+\mu-{\xi}^{2}/2
\right]^{2}\rd t+\frac1{2}\lambda\int_0^{\tau} \re^{-\tilde{y}}\rd t,\notag\\
\tilde{\sigma}^2&=(1-\rho^2)\frac1{\tau}\int_0^{\tau} \re^{\tilde{y}}\rd t,\notag\\
\tilde{r}&=r-\frac1{2}\rho^2\frac1{\tau}\int_0^{\tau}
\re^{\tilde{y}}\rd t+
\rho\left(\frac{\xi}{4}-\frac{\mu}{\xi}\right)\frac1{\tau}\int_0^{\tau}
\re^{\tilde{y}/2}\rd t-\frac{\rho}{\xi}\frac1{\tau}\int_0^{\tau}
\re^{\tilde{y}/2}v(t)\rd t
-\frac{\lambda\rho}{\xi}\frac1{\tau}\int_0^{\tau}
\re^{-\tilde{y}/2}\rd t.
\end{align}
Performing the inverse Fourier transform we obtain the evolution operator
kernel:
\begin{align}\label{Kern2:MG}
g(\tau,x-x_0\,,y,y_0)&=\bar{\mathcal{N}}(\tau)\!\!\int\mathcal{D}v
 \re^{S_0}\frac{ \re^{-\tau
r}}{\sqrt{2\pi\tau\tilde{\sigma}^2}}\times\notag\\
&\times\exp\left(-\frac1{2}
\frac{\left(x-x_0+\tau\left[\tilde{r}-{\tilde{\sigma}^2}/{2}\right]\right)^2}{\tau\tilde{\sigma}^2}\right)
\delta\left(y-y_0-\int_0^\tau v(t)\rd t\right).
\end{align}
Thus, formula~\eqref{Kern2:MG} is a path integral representation
of the evolution operator kernel for the Merton-Garman
Hamiltonian in the case $\alpha=1$. As it was mentioned earlier,
the comparison of this result with the formula
from~\cite{Baaq,Baaq97} shows that the latter does not contain the
second summand in $S_0$\,, as well as there is a change in the
corresponding numerical coefficient in the third summand
$\tilde{r}$. Recall that in~\cite{Baaq,Baaq97} there is used a
Feynman path integral defined in the configurational phase space,
in contrast to the kernel~\eqref{Kern2:MG} which is represented in
the velocity phase space. The connection between both
representations can be found in appendix~\ref{Path:y_v}.

Making use of this kernel, we obtain a generalization of the
Black-Scholes result~\eqref{price:BS} for the option price. For
this to be done, we need to multiply the kernel
${g(\tau,x-x_0\,,y,y_0)}$ by the payment function~\eqref{prifun}
and integrate over the variables $x_0,\, y_0$. Since function
\eqref{prifun} does not depend on $y_0$\,, the integral with
respect to $y_{0}$ is easily performed owing to the presence of
$\delta$-function. As a result we obtain:
\begin{align}\label{price:MG}
C(t,S(t))= \bar{\mathcal{N}}(\tau)\!\!\int\mathcal{D}v
 \re^{S_0}\left[S(t) \re^{\tau{\Delta r}}N(\tilde{d}_+)- \re^{-\tau
r}KN(\tilde{d}_{-})\right],
\end{align}
where we denoted $\Delta r=\tilde{r}-r$ and quantities
$\tilde{d}_\pm$ are determined by~\eqref{price:BS} for $d_\pm$ by
means of the change $r\rightarrow \tilde{r}$, $\sigma\rightarrow
\tilde{\sigma}$.

\subsection{The case of arbitrary $\alpha$-values}\label{alphaD}

In this case, in order to obtain the path integral representation
of the evolution operator we will act the following way. First, we
will make use of the path integral~\eqref{gaus:1} for
linearization of the third summand in~\eqref{HmF:MG}:
\begin{align}\label{EvOpA1:MG}
 \re^{-\tau \tilde{H}_{\rm MG}(k)}&=\,{\mathcal{N}}(\tau)\!\!\int\mathcal{D}w
\exp\left(-\frac1{2}\int_0^\tau w^2(t) \rd
t\right)\times\notag\\&\times T\exp\left(-\int_0^\tau h_0(k,y) \rd
t-\int_0^\tau \left[h_1(k,y)+\xi w(t)
\re^{y(\alpha-1)}\right]\frac{\partial}{\partial y}\, \rd
t\right).
\end{align}
Then, we use the path integral~\eqref{gaus:2} for linearization of
the summand with the derivation operator~\eqref{EvOpA1:MG}.
Preliminarily, let us denote:
\begin{align*}
&A(t)=-\frac{1}{2}\left[h_1(k,y)+\xi w(t) \re^{y(\alpha-1)}\right],\qquad B(t)=\frac{\partial}{\partial y}\,,\\
&[A(t),B(t)]_{-}=\frac{1}{2}\left[h'_1(k,y)+(\alpha-1)\xi w(t)
\re^{y(\alpha-1)}\right]=-A'(t).
\end{align*}
The sense of the previous transformations, as we observed before,
is in splitting a shift-operator which permits to perform the
$T$-operation. As a result, for the $T$-exponent~\eqref{EvOpA1:MG}
we obtain the relationship:
\begin{align}\label{Texp:MG}
&T\exp\left(-\int_0^\tau h_0(k,y) \rd t-\int_0^\tau
\left[h_1(k,y)+\xi w(t)
\re^{y(\alpha-1)}\right]\frac{\partial}{\partial y}\, \rd
t\right)=\notag\\&={{\mathcal{N}}(\tau)}^2\!\!\int\mathcal{D}v\mathcal{D}u\,
\exp\left(-\frac1{2}\int_0^\tau v^2(t) \rd t-\ri\int_0^\tau
v(t)u(t) \rd t\right)\times\notag\\&\times T\exp\left(-\int_0^\tau
\left\{\left[v(t)+2\,\ri
u(t)\right]A(t)+v(t)\frac{\partial}{\partial
y}+A'(t)+h_0(k,y)\right\} \rd t\right).
\end{align}
Disentangling the exponent containing the shift-operator and
substituting it into~\eqref{EvOpA1:MG}, we write down the
evolution operator in the form:
\begin{align}\label{Texp2a:MG}
& \re^{-\tau \tilde{H}_{\rm
MG}(k)}={{\mathcal{N}}(\tau)}\!\!\int\mathcal{D}v\,\exp\left(-\frac1{2}\int_0^\tau
v^2(t) \rd t\right)\Phi(v)\exp\left(-\int_0^\tau h_0(k,\tilde{y})
\rd t\right)\exp\left(-\int_0^\tau v(t) \rd
t\frac{\partial}{\partial y}\right),
\end{align}
where we denoted:
\begin{align}\label{Texp2:MG}
\Phi(v)&=\,{\mathcal{N}}(\tau)\!\!\int\mathcal{D}u\exp\left(-\ri\int_0^\tau
v(t)u(t) \rd
t\right){\mathcal{N}}(\tau)\!\!\int\mathcal{D}w\,\exp\left(-\frac1{2}\int_0^\tau
w^2(t) \rd t\right)\times\notag\\&\times\exp\left(-\int_0^\tau
\left\{[v(t)+2\ri u(t)]\tilde{A}(t)+\tilde{A}'(t)\right\} \rd
t\right).
\end{align}
Let us note here that (\verb#~#)-barred quantities differ from
those, used before, by the change to $y\rightarrow \tilde{y}$,
where $\tilde{y}=y-\int_t^{\tau}v(t') \rd t'$.

While calculating $\Phi(v)$ it is necessary to perform integration with respect
to variables $w(t)$ and $u(t)$. Since the mentioned integrals are Gaussian,
they can be performed in closed form. Upon simple but a bit cumbersome
calculations, which are omitted, we obtain:
\begin{align}\label{EvOpA2:MG}
& \re^{-\tau
\tilde{H}(k)}=\bar{\mathcal{N}}(\tau)\!\!\int\mathcal{D}v\prod_t
\re^{-\tilde{y}(\alpha-1)} \exp\left(-\frac1{2\xi^2}\int_0^\tau
\frac{\left[v(t)-h_1(k,\tilde{y})\right]^2}{
\re^{2\tilde{y}(\alpha-1)}} \rd
t\right)\times\notag\\&\times\exp\!\left({-\int_0^\tau
h_0(k,\tilde{y}) \rd t+\frac1{2}\int_0^\tau h'_1(k,\tilde{y}) \rd
t+\frac1{2}(\alpha-1)\int_0^\tau
\left[v(t)-h_1(k,\tilde{y})\right] \rd
t}\right)\exp\left(-\int_0^\tau v(t) \rd t\frac{\partial}{\partial
y}\right).
\end{align}
Respectively, for the Fourier representation of the
kernel~\eqref{Kern0:MGk} one obtains the relationship:
\begin{align}\label{Kern:MG_alp}
\tilde{g}(\tau,k,y,y_0)=&\bar{\mathcal{N}}(\tau)\!\!\int\mathcal{D}v\prod_t
\re^{-\tilde{y}(\alpha-1)} \exp\left(-\int_0^\tau
L(k,v(t),\tilde{y})\, \rd t\right)\delta\left(y-y_0-\int_0^\tau
v(t) \rd t\right),
\end{align}
where we denoted the function:
\begin{align}\label{Lagr}
L(k,v,y)=\frac1{2\xi^2}\frac{\left[v-h_1(k,y)\right]^2}{
\re^{2y(\alpha-1)}}+h_0(k,y)
-\frac1{2}h'_1(k,y)+\frac1{2}(\alpha-1)\left[v-h_1(k,y)\right].
\end{align}
Let us mention here that the representation ~\eqref{Kern:MG_alp}
corresponds to the structure~\cite{Klein, Blaz, Grosch} of the
path integrals subject to the curvilinear coordinates. As is seen
from~\eqref{Kern:MG_alp} the kernel is not determined by means of
a ``classical'' Lagrange function which is assigned to the
Hamiltonian~\eqref{Hm:MG}. It is not hard to observe that this
Lagrange function (for the imaginary time) is determined by the
relationship:
\begin{align}\label{Lagr:0}
L_0(k,v,y)=\frac1{2\xi^2}\frac{\left[v-\bar{h}(k,y)\right]^2}{
\re^{2y(\alpha-1)}}+h_0(k,y),
\end{align}
where we denoted
\begin{align}\label{par_ha}
\bar{h}(k,y)&=-\ri\rho\xi \re^{y(\alpha-1/2)}k-\left(\lambda
 \re^{-y}+\mu-\frac{\xi^2}{2} \re^{2y(\alpha-1)}\right),
\end{align}
the quantity  $h_0(k,y)$ is determined by ~\eqref{par_h}. Note
also that $\bar{h}(k,y)$ is different from $h_1(k,y)$
\eqref{par_hhhh} by a factor $\alpha$ in the last summand. It is
evident that the difference between $L(k,v,y)$ and $L_0(k,v,y)$
determines the above mentioned additional terms. The corresponding
terms are not taken into account in~\cite{Baaq} for the
constructed evolution operator kernel of the Hamiltonian
\eqref{Hm:MG}. Note also that at the change $L(k,v,y)\rightarrow
L_0(k,v,y)$ in~\eqref{Kern:MG_alp}, the obtained kernel will
correspond not to operator~\eqref{Hm:MG}, but to its analog
$H_{\rm MG}^*$\,, being in some way symmetrized over operators
${\partial}/{\partial y}$. The structure $H_{\rm MG}^*$ can be
determined, in particular, making use of the approach devised in
\cite{Blaz}.

The next step, having split the dependence on the $k$-variable, one can obtain:
\begin{align}\label{Kern11:MG}
\tilde{g}(\tau,k,y,y_0)&= \re^{-\tau
r}\bar{\mathcal{N}}(\tau)\!\!\int\mathcal{D}v\prod_t
\re^{-\tilde{y}(\alpha-1)}
 \re^{S_0}\times\notag\\
&\times \exp\left(-\frac1{2}\tau\tilde{\sigma}^2k^2-\ri\tau
k\left[{\tilde{\sigma}^2}/2-\tilde{r}\right]\right)\delta\left(y-y_0-\int_0^\tau
v(t) \rd t\right),
\end{align}
where we denoted:
\begin{align}\label{Par2}
S_0=&-\frac1{2\xi^2}\int_0^{\tau}\frac{\bigl[v(t)-h(\tilde{y})\bigr]^2}{
\re^{2\tilde{y}(\alpha-1)}} \rd t+
\frac{1}{2}\int_0^{\tau}h'(\tilde{y}) \rd t+\frac1{2}(\alpha-1)\int_0^{\tau}\bigl[v(t)-h(\tilde{y})\bigr] \rd t,\notag\\
\tilde{\sigma}^2=&\,(1-\rho^2)\frac1{\tau}\int_0^{\tau} \re^{\tilde{y}} \rd t,\notag\\
\tilde{r}=\,&r-\rho^2\frac1{\tau}\int_0^{\tau} \re^{\tilde{y}} \rd
t-\frac{\rho\xi}{4}\frac1{\tau}\int_0^{\tau}
\re^{\tilde{y}(\alpha-1/2)} \rd t
-\frac{\rho}{\xi}\frac1{\tau}\int_0^{\tau}\left[v(t)-h(\tilde{y})\right]
\re^{\tilde{y}(3/2-\alpha)} \rd t,\notag\\
h(y)=&-\left(\lambda \re^{-y}+\mu-\frac{\xi^2\alpha}{2}
\re^{2y(\alpha-1)}\right).
\end{align}
Here $h'(y)$ means the derivative of function with respect to a suitable
argument. Performing the inverse Fourier transform we obtain the evolution
operator kernel in the case of arbitrary $\alpha$-values.
\begin{align}\label{Kern2:MG_Al}
g(\tau,x-x_0\,,y,y_0)=&\,\bar{\mathcal{N}}(\tau)\!\!\int\mathcal{D}v
\prod_t \re^{-\tilde{y}(\alpha-1)} \re^{S_0}\frac{ \re^{-\tau
r}}{\sqrt{2\pi\tau\tilde{\sigma}^2}}
\times\notag\\\times&\exp\left(-\frac1{2}\frac{\bigl(x-x_0
+\tau\left[\tilde{r}-{\tilde{\sigma}^2}/{2}\right]\bigr)^2}
{\tau\tilde{\sigma}^2}\right)\delta\left(y-y_0-\int_0^\tau v(t)
\rd t\right).
\end{align}

Then, the generalization of the Black-Scholes for the option
price will take the form:
\begin{align}\label{priceA:MG}
C(t,S(t))=\bar{\mathcal{N}}(\tau)\!\!\int\mathcal{D}v\prod_t
\re^{-\tilde{y}(\alpha-1)} \re^{S_0}\left[S(t)  \re^{\tau{\Delta
r}}N(\tilde{d}_+)- \re^{-\tau r}KN(\tilde{d}_{-})\right].
\end{align}

\section{Some comments concerning path integral calculations}

Works~\cite{Baaq,Baaq97,Hull} present the formulas for the kernel
calculation~\eqref{kernF:y_0} in the case $\alpha=1$, $\lambda=0$.
Meanwhile, the sub-integrand in~\eqref{kernF:y_0} was represented
as a three-tuple series:
\begin{align}\label{kernF:S1}
&\frac1{\sqrt{2\pi\tau\tilde{\sigma}^2}}\exp\left(-\frac1{2}\frac{\left(x-x_0+
\tau\left[\tilde{r}-{\tilde{\sigma}^2}/{2}\right]\right)^2}{\tau\tilde{\sigma}^2}\right)\rightarrow\notag\\
&\rightarrow\sum_{n,m,k}a_{n\,m\,k}\left(\frac1{\tau}\int_0^{\tau}
\re^{y(t)} \rd t\right)^n \left(\frac1{\tau}\int_0^{\tau}
\re^{y(t)/2} \rd t\right)^m\left( \re^{y(0)/2}\right)^k.
\end{align}
Then, for a general series term:
\begin{align*}
\int_{-\infty}^{\infty}\rd y_0\int_{y_0}^y\mathcal{D'}y
 \re^{S_0}\left(\frac1{\tau}\int_0^{\tau} \re^{y(t)} \rd t\right)^n
\left(\frac1{\tau}\int_0^{\tau} \re^{y(t)/2} \rd t\right)^m\left(
\re^{y(0)/2}\right)^k
\end{align*}
there was obtained its closed form. Nonetheless one can be easily
convinced that such an approach to calculations is wrong, since
the used series expansion~\eqref{kernF:S1} does not exist.

Thus, it is reasonable to mention some approximate methods of
calculating the path integrals, in particular, for the case
$\alpha=1$. Consider a formula for the option
price~\eqref{price:MG}. We will represent the corresponding path
integrals in~\eqref{price:MG}  as those with respect to the Gauss
measure:
\begin{align}\label{price:gauss}
C(t,S(t))=\int \rd \mu(v)\left[S(t)F_{+}(v)- \re^{-\tau
r}K\,F_{-}(v)\right],
\end{align}
where $\rd\mu(v)=\bar{\mathcal{N}}(\tau)\mathcal{D}v \re^{S_{\rm
G}}$~-- means a probabilistic Gauss integration measure. Here, the
quantity $S_{\rm G}$ is given by the expression:
\begin{align*}
S_{\rm G}&=-\frac1{2\xi^2}\int_0^{\tau}\left[ v(t)+\mu-{\xi}^{2}/2
\right]^{2} \rd t,
\end{align*}
and the next terms $F_{\pm}(v)$ hold the remaining factors of
formula \eqref{price:MG}. It is easy to see that the values
$F_{\pm}(v)$ are bound by $v\in\mathbb{R}$. Therefore, the
integrals used in~\eqref{price:gauss} exist. For an approximate
calculation of path integrals~\eqref{price:gauss} one can use
numerical quadrature formulas, presented, in particular,
in~\cite{Jydkov_99}.

The use of approximation approaches is also of some interest. In
solving the problems of quantum mechanics and statistical physics,
the mean-path-approximation~\cite{Feyn_Hib, Feyn} was effectively
used. From this point of view let us rewrite formula
\eqref{price:MG} in the form:
\begin{align*}
C(t,S(t))= \bar{\mathcal{N}}(\tau)\!\!\int\!\mathcal{D}v\left[S(t)
\re^{S_{+}}- \re^{-\tau r}K \re^{S_{-}}\right].
\end{align*}
The mean-path approximations  could be naturally considered as
zero-approximations. In this case within the  quantities $S_{\pm}$ we will make
the change:
\begin{align*}
\tilde{y}(t)\rightarrow y_{m}=
\frac1{\tau}\int_0^{\tau}{\tilde{y}(t)} \rd t=y
-\frac1{\tau}\int_0^{\tau}t\,v(t) \rd t.
\end{align*}
Upon the  above mentioned change,  the quantities $S_{\pm}$ will depend on the
integration variables via the expressions:
\begin{align*}
\int_0^{\tau}v(t) \rd t,\qquad \frac1{\tau}\int_0^{\tau}t\,v(t)
\rd t
\end{align*}
and the path integrals in~\eqref{price:MG} will then reduce to the
usual quadratures. As a result, the price option formula will take
the form:
\begin{align}\label{price0:MG}
C(t,S(t))= \frac{2\sqrt{3}}{\xi^2\tau}
\re^{-\tau^2(\mu-\xi^2/2)/2}\int \frac{\rd v_1\rd v_2}{2\pi}
\re^{S^*_0}\left[S(t) \re^{\tau{\Delta r^*}}N(d^*_+)- \re^{-\tau
r}KN(d^*_{-})\right],
\end{align}
where the corresponding quantities are defined by the expressions:
\begin{align*}
S^*_0=&-\frac1{\xi^2\tau}\left(2v_1^2+6v_2^2-6v_1v_2-\tau
v_1\left[\mu-\xi^2/2\right]\right)-\frac{\lambda\tau}{\xi^2} \re^{-y+v_2}(\mu-\xi^2/2)-\\
&-\frac{\lambda^2\tau}{2\xi^2}
\re^{-2(y-v_2)}+\frac{\lambda}{\xi^2} \re^{-y}(1- \re^{v_1})+
\frac{\lambda\tau}2  \re^{-y+v_2},\\
d^*_\pm=&\,\frac{\ln\left[S(t)/K\right]+\tau\left(r^*\pm{{\sigma^*}^2}/2\right)}{\sigma^*\sqrt{\tau}},\\
\sigma^*=&\,\left(1-\rho^2\right) \re^{y-v_2},\\
r^*=&\,r-\frac1{2}\rho^2
\re^{y-v_2}+\rho\left(\frac{\xi}4-\frac{\mu}{\xi}\right)
\re^{(y-v_2)/2}-
\frac{2\rho}{\xi} \re^{{y}/2}(1- \re^{-v_1}),\\
\Delta r^*=&\,r^*-r.
\end{align*}
Our next step is to take the series corrections into account, if we expand the
quantities $S_{\pm}$ into series with respect to suitable deviations
$\Delta y(t)$:
\begin{align*}
\tilde{y}(t)=y_{m}+\Delta y(t)\,.
\end{align*}
Certainly, the approaches mentioned above need a special
investigation.

\section{Conclusions}

Using the Gaussian path integral, a representation of the
evolution operator kernel for the Merton-Garman model was
obtained. Also, some shortcomings that occurred in the cited
sources~\cite{Baaq97, Baaq} were explained and corrected. A
generalization of the well-known Black-Scholes formula for the
rational price of European call option was also presented.
Approximate approaches to the calculation of path integrals were
also described.

Note also that the Black-Scholes formula~\eqref{price:BS} is
widely used in practice in analyzing the options~\cite{Hull_03}.
It is known that the analyzed model gives rise to quite good
results in the case of well developed markets. However,
shortcomings occur when a stock market is in an uncertainty state
and the price dynamics swings~\cite{Nazar_07}. At the same time,
the Merton-Garman model should be considered as more realistic
generalization of the Black-Scholes model taking into account the
arbitrary volatility nature (formulas~\eqref{V}, \eqref{dyn:SV}).
That is why the analysis of the obtained~\eqref{price:MG}
and~\eqref{priceA:MG} formulas presents a considerable interest
which will be the subject of our next investigations.

\section*{Acknowledgments} One of the authors (V. Yanishevsky) is especially
grateful to A.K.~Prykarpatsky for the fruitful discussions and
the valuable comments and remarks.

\appendix

\section{The Black-Scholes model}\label{Bl_Sch}

\subsection{The Black-Scholes equation}

The Black-Scholes equation~\cite{Schyr,Baaq} describes the $C(t)$
option price dynamics:
\begin{align}\label{eq:BS}
    \frac{\partial C(t)}{\partial t}=H_{\rm BS}C(t),
\end{align}
where $H_{\rm BS}$ denotes the following Black-Scholes
Hamiltonian:
\begin{align}\label{hm:BS}
H_{\rm BS}=-\frac{\sigma^2}2\frac{\partial^2}{\partial
x^2}+\left(\frac{\sigma^2}2-r\right)\frac{\partial}{\partial x}+r.
\end{align}
Equation~\eqref{eq:BS} visually resembles a Schrodinger equation,
written with respect to imaginary time, or a diffusion equation.
Quantities $r $ and $\sigma $ are considered to be constant. The
general solution of equation~\eqref{eq:BS} is represented in the
form:
\begin{align}\label{sol:BS}
C(t)= \re^{tH_{\rm BS}}C(0),
\end{align}
where $C(t)$ is the value in time $t$, $C(0)$ is an initial
condition. When using equation~\eqref{eq:BS} in order to evaluate
the option prices it is convenient to study the dynamics backward
in time. For a given value of $C(T)$ at the final moment of time
$T$ we need to evaluate $C(t)$ at $t<T$. It follows from the
solution of equation~\eqref{sol:BS} that
\begin{align}\label{sol:BS2}
C(t)= \re^{-(T-t)H_{\rm BS}}C(T)= \re^{-\tau H_{\rm BS}}C(T),
\end{align}
where we denoted $\tau =T-t$. Making use of the kernel $g(\tau
,x-x_{0}),$ for expression~\eqref{sol:BS2} one obtains that:
\begin{align}\label{ker:BS}
C(t,x)=\int_{-\infty}^{\infty}g(\tau,x-x_0)C(T,x_0)\rd x_0\,.
\end{align}
Based on~\eqref{sol:BS2} and~\eqref{ker:BS} one finds the
relationship:
\begin{align}\label{ker:BS_2}
g(\tau,x-x_0)= \re^{-\tau H_{\rm BS}}\delta(x-x_0).
\end{align}
Since Hamiltonian~\eqref{hm:BS} coefficients are constants, it is
convenient to proceed to Fourier transform by the formula:
\begin{align*}\label{eq2:BS}
    f(t,x)=\frac1{\sqrt{2\pi}}\int_{-\infty}^{\infty}\tilde{f}(t,k) \re^{\ri kx}\rd
    k.
\end{align*}
In equation~\eqref{eq:BS} it is convenient to proceed to Fourier
transform, receiving the following result:
\begin{equation}\label{eqf:BS2}
    \frac{\partial \tilde{C}(t,k)}{\partial t}=\tilde{H}_{\rm BS}(k)\tilde{C}(t,k),
\end{equation}
where $\tilde{C}(t,k)$ denotes Fourier transform of proper
function and Fourier transform of Hamiltonian~\eqref{hm:BS}:
\begin{equation}\label{hmf:BS}
\tilde{H}_{\rm
BS}(k)=\frac{\sigma^2}2k^2+\ri\left(\frac{\sigma^2}2-r\right)k+r\,.
\end{equation}
An equation~\eqref{eqf:BS2} for a Fourier transform will be of the
form:
\begin{align}\label{solF:BS2}
\tilde{C}(t,k)= \re^{-\tau \tilde{H}_{\rm BS}(k)}\tilde{C}(T,k)\,.
\end{align}
Respectively for the kernel Fourier image  $g(\tau,x-x_0)$ one has:
\begin{align*}
\tilde{g}(\tau,k)= \re^{-\tau \tilde{H}_{\rm BS}(k)}.
\end{align*}
Similarly, for the kernel of the evolution
operator~\eqref{eqf:BS2}, applying inverse Fourier transform, we
receive the following expression:
\begin{align}\label{ker2:BS}
g(\tau,x-x_0)&=\frac{ \re^{-\tau
r}}{\sqrt{2\pi}}\int_{-\infty}^{\infty}\tilde{g}(\tau,k) \re^{\ri
k(x-x_0)}\rd k=\notag\\&= \frac{ \re^{-\tau
r}}{\sqrt{2\pi}}\int_{-\infty}^{\infty}\exp\left(-\frac1{2}\tau\sigma^2k^2-\ri\tau
k\left[{\sigma^2}/2-r\right]\right) \re^{\ri k(x-x_0)}\rd k=\notag\\
&=\frac{ \re^{-\tau
r}}{\sqrt{2\pi\tau\sigma^2}}\exp\left(-\frac1{2}\frac{\left(x-x_0+\tau
\left[r-{\sigma^2}/{2}\right]\right)^2}{\tau\sigma^2}\right).
\end{align}

\subsection{A Black-Scholes formula}

Let us recall a classical example of evaluation for the rational
price of the European call option~\cite{Schyr, Baaq}. A boundary
condition (a payoff function) is defined in the form:
\begin{align}\label{prifun}
C(T,S)=(S-K)^+=\left\{%
\begin{array}{ll}
    S-K, & S>K, \\
    0, & S\leqslant K, \\
\end{array}%
\right.
\end{align}
where $K$ is strike price, $S$ is a market price at a moment $T$
(time to maturity). The payoff function~\eqref{prifun} has a
sufficiently obvious meaning~-- option transaction realizes if the
stock market price is greater than the strike price. Option price
at time $t<T$ is evaluated making use of the formula:
\begin{align}\label{price}
C(t,S(t))&=\int_{-\infty}^{\infty}g\left(\ln\left[S(t)\right]-x_0\right)
\left( \re^{x_0}-K\right)^+\rd x_0=\notag\\
&=\frac{ \re^{-r\tau}}{\sqrt{2\pi\sigma^2}}\int_{\ln(K)}^{\infty}
\exp\left[-\frac1{2}\frac{\left(\ln[S(t)]-x_0+\tau[r-{\sigma^2}/{2}]\right)^2}{\tau\sigma^2}\right]
( \re^{x_0}-K)\rd x_0\,,
\end{align}
where $S(t)$ denotes stock price at time $t$.\par Having completed
the necessary computations we receive, for option
price~\cite{Black}, the well known Black-Scholes formula:
\begin{align}\label{price:BS}
C(t,S(t))=S(t)N(d_+)-K \re^{-r\tau}N(d_{-}).
\end{align}
We have here denoted:
\begin{align*}
N(x)=\frac1{2\pi}\int_{-\infty}^x
 \re^{-z^2/2}\rd z,\qquad d_\pm=\frac{\ln[S(t)/K]+\tau\left(r\pm{\sigma^2}/2\right)}
 {\sigma\sqrt{\tau}}\,.
\end{align*}

\section{Gaussian path integrals}\label{Tprod}

The Gaussian path integral is used for the degree linearization in
an $T$-exponential and looks as follows:
\begin{align}\label{gaus:1}
T&\exp\left(\int_0^{\tau}\frac1{2}A^2(t) \rd t+\int_0^{\tau}B(t)d t\right)=\notag \\
=&\,{\mathcal{N}}(\tau)\!\!\int\mathcal{D}v
\exp\left(-\frac1{2}\int_0^\tau v^2(t) \rd
t\right)T\exp\left(-\int_0^\tau v(t)A(t) \rd t+\int_0^\tau B(t)
\rd t\right).
\end{align}
In this work $A(t)$ and $B(t)$ -- differential operators (in general not
commutative), ``the constant'' ${\mathcal{N}}(\tau)$ is determined from the
condition:
\begin{align*}
{\mathcal{N}}(\tau)\!\!\int\mathcal{D}v
\exp\left(-\frac1{2}\int_0^\tau v^2(t) \rd t\right)=1,\qquad
\mathcal{D}v=\prod_{0\leqslant t\leqslant \tau} \rd v(t)\,.
\end{align*}
The Gaussian path integral~\eqref{gaus:1} was used in~\cite{Blaz}
in constructing the path integral Green function representation
for some quantum mechanical Hamiltonians.
\par It is evident that the field of application of the formula~\eqref{gaus:1} is wide enough.
The path integral~\eqref{gaus:1} is interpreted as an operator
identity. In particular, one can ``disentangle'' the factor
$T\exp\left(\int_0^{\tau}B(t) \rd t\right)$ in~\eqref{gaus:1} in
accordance with~\eqref{Texp}, and obtain the equivalent equality:
\begin{align}\label{gaus:1a}
T\exp\left(\frac1{2}\int_0^{\tau}\tilde{A}^2(t)\rd t\right)=
{\mathcal{N}}(\tau)\!\!\int\mathcal{D}v
\exp\left(-\frac1{2}\int_0^\tau v^2(t) \rd
t\right)T\exp\left(-\int_0^\tau v(t)\tilde{A}(t) \rd t\right),
\end{align}
where
\begin{align*}
\tilde{A}(t)=\left[T\exp\left(\int_0^{t}B(t') \rd
t'\right)\right]^{-1}A(t)\,T\exp\left(\int_0^{t}B(t') \rd
t'\right).
\end{align*}
The integration in~\eqref{gaus:1a} is realized with respect to the
Gaussian probability measure, since:
\begin{align*}
\mathcal{N}(\tau)\!\!\int\mathcal{D}v
\exp\left(-\frac1{2}\int_0^\tau v^2(t) \rd
t\right)T\exp\left(-\int_0^\tau v(t)\tilde{A}(t) \rd t\right)=\int
\rd\mu(v)T\exp\left(-\int_0^\tau v(t)\tilde{A}(t) \rd t\right),
\end{align*}
where $\rd\mu(v)=\mathcal{N}(\tau)\mathcal{D}v
\exp\left(-\frac1{2}\int_0^\tau v^2(t) \rd t\right)$.

\par To check the identity~\eqref{gaus:1a}, we expand the left and
right hand sides in degree series. In particular, $T$-exponent on
the left hand side~\eqref{gaus:1a} is represented as the degree
series
\begin{align}\label{Texp:1}
T\exp\left(\frac1{2}\int_0^{\tau}\tilde{A}^2(t)\rd
t\right)=\sum_{n=0}^{\infty}\frac{1}{n!}T\left(\frac1{2}\int_0^{\tau}\tilde{A}^2(t)\rd
t\right)^n.
\end{align}
Expanding the right hand side~\eqref{gaus:1a}
$T\exp\left(-\int_0^\tau v(t)\tilde{A}(t) \rd t\right)$ in series
and using the Gauss measure properties, one can derive  expression
\eqref{Texp:1}. Really, for the left hand side of equality
\eqref{gaus:1a} one has:
\begin{align}\label{Texp:2}
\int \rd\mu(v)T\exp\left(-\int_0^\tau v(t)\tilde{A}(t) \rd
t\right)=\sum_{n=0}^{\infty}\int_0^\tau  \rd t_1\ldots\int_0^\tau
 \rd
 t_n\frac{(-1)^n}{n!}\mu_nT\left[\tilde{A}(t_1)\ldots\tilde{A}(t_n)\right],
\end{align}
where $\mu_n$ is the $n$-th Gauss measure correlation moment. As is well known,
all uneven moments vanish, and the even ones are expressed by means of
the second correlation moment:
\begin{align*}
\mu_2=\int \rd\mu(v)v(t_1)v(t_2)=\delta(t_1-t_2).
\end{align*}
The moment of  $2\,n$-th order is expressed by means of the moment $\mu_2$ as
the relationship:
\begin{align}\label{mu_2n}
\mu_{2n}=\int \rd\mu(v)v(t_1)\ldots
v(t_{2n})=\sum_{P}\mu_2^{1}\ldots\mu_2^{n}\,.
\end{align}
A summand in~\eqref{mu_2n} is taken over all possible different
smashes of $2n$ elements by two-element subsets (see, for
instance,~\cite{Zin_Ju}). The amount of summands in~\eqref{mu_2n}
equals
\begin{align*} \frac{(2n)!}{2^nn!}\,.
\end{align*}
Making use of symmetry properties (the permutations of operators
under the $T$-product sign) it is not hard to demonstrate that all
these summands give rise to the same result, thereby the
formula~\eqref{Texp:2} reduces to~\eqref{Texp:1}.
\par Based on~\eqref{gaus:1} and the operator identity:
\begin{align*}
2A(t)B(t)=\frac1{2}(A(t)+B(t))^2-\frac1{2}(A(t)-B(t))^2+[A(t),B(t)]_{-}\,,
\end{align*}
let us linearize the product of operators $A(t)B(t)$:
\begin{align}\label{gaus:2}
&T\exp\left(\int_0^{\tau}\left[2A(t)B(t)+D(t)\right]\rd t\right)=
{\mathcal{N}}(\tau)^2\!\!\int\mathcal{D}v\mathcal{D}u\,
\exp\left(-\frac1{2}\int_0^\tau v^2(t) \rd t-\frac1{2}\int_0^\tau
u^2(t) \rd t\right)\times\notag\\&\times T\exp\left(-\int_0^\tau
\Bigl\{[v(t)+\ri u(t)]A(t)+[v(t)-\ri
u(t)]B(t)-[A(t),B(t)]_{-}-D(t)\Bigr\} \rd t\right).
\end{align}
Here we have used the following notation of the operator
commutator $[A(t),B(t)]_{-}=A(t)B(t)-B(t)A(t)$. Using the shift
$v(t)\rightarrow v(t)+\ri u(t)$ the right hand side of
equation~\eqref{gaus:2} \ can be written in the form:
\begin{align}\label{gaus:3}
&{\mathcal{N}}(\tau)^2\!\!\int\mathcal{D}v\mathcal{D}u\,
\exp\left(-\frac1{2}\int_0^\tau v^2(t) \rd t-\ri\int_0^\tau
v(t)u(t) \rd t\right)\times\notag\\&\times T\exp\left(-\int_0^\tau
\Bigl\{[v(t)+2\,\ri u(t)]A(t)+v(t)B(t)-[A(t),B(t)]_{-}-D(t)\Bigr\}
\rd t\right).
\end{align}
In the case of differential operators, the relationship
\eqref{gaus:2} makes it possible to decrease its differential
order. Note also that the path integral representation
\eqref{gaus:2} is equivalent to the Stratonovich-Hubbard
identity~\cite{Masu} for the density matrix in quantum statistics
problems.
\par Jointly with the introduced formulas, the following expression is used to extract
$T$-exponent operators~\cite{Nazaj}:
\begin{align}\label{Texp}
T\exp\left(\int_0^{\tau}[A(t)+B(t)]\rd
t\right)=T\exp\left(\int_0^{\tau}A(t) \rd t\right)
T\exp\left(\int_0^{\tau}\tilde{B}(t)\rd t\right),
\end{align}
where
\begin{align*}
\tilde{B}(t)=\left[T\exp\left(\int_0^{t}A(t') \rd
t'\right)\right]^{-1}B(t)\,T\exp\left(\int_0^{t}A(t') \rd
t'\right).
\end{align*}

\section{A connection between path integral approaches}\label{Path:y_v}

Consider the case for parameter values $\alpha=1,\,\lambda=0$. Based on
\eqref{Kern2:MG} one can write down the kernel expression as:
\begin{align}\label{Kern2:MG_y}
g(\tau,x-x_0\,,y,y_0)=\int_{y_0}^{y}\mathcal{D'}y \re^{S_0}\frac{
\re^{-\tau r}}{\sqrt{2\pi\tau\tilde{\sigma}^2}}\exp
\left(-\frac1{2}\frac{\left(x-x_0+\tau\left[\tilde{r}-{\tilde{\sigma}^2}/{2}\right]\right)^2}
{\tau\tilde{\sigma}^2}\right).
\end{align}
The measure of integration in~\eqref{Kern2:MG_y} satisfies the
condition:
\begin{align*}
\int_{y_0}^{y}\mathcal{D'}y \exp\left[-\frac1{2\xi^2}\int_0^{\tau}
\dot{y}(t)^2 \rd
t\right]_{\tau\rightarrow0}\rightarrow\delta(y-y_0),
\end{align*}
at the boundary values $y(0)=y_{0}\,,\,\,y(\tau )=y$. Moreover, in
the above expressions \eqref{Par1}  one should make the following
substitutions:
\[\tilde{y}\rightarrow y(t),\qquad v(t)\rightarrow \dot{y}(t).\]
Thereby, we obtain:
\begin{align}\label{kernF:y_0}
\tilde{g}(\tau,x-x_0\,,y)=\int_{-\infty}^{\infty}\rd
y_0\int_{y_0}^{y}\mathcal{D'}y
 \re^{S_0}\frac{ \re^{-\tau
r}}{\sqrt{2\pi\tau\tilde{\sigma}^2}}
\exp\left(-\frac1{2}\frac{\left(x-x_0+\tau\left[\tilde{r}-{\tilde{\sigma}^2}/{2}\right]\right)^2}{\tau\tilde{\sigma}^2}\right),
\end{align}
where
\begin{align}\label{ParF:y0}
S_0&=-\frac1{2\xi^2}\int_0^{\tau}\left[ \dot{y}(t)+\mu-{\xi}^{2}/2
\right]^{2} \rd t,\notag\\
\tilde{\sigma}^2&=(1-\rho^2)\frac1{\tau}\int_0^{\tau} \re^{y(t)} \rd t,\notag\\
\tilde{r}&=r-\frac1{2}\rho^2\frac1{\tau}\int_0^{\tau} \re^{y(t)}
\rd t+
\rho\left(\frac{\xi}{4}-\frac{\mu}{\xi}\right)\frac1{\tau}\int_0^{\tau}
\re^{y(t)/2} \rd t -\frac{2\rho}{\xi}\frac1{\tau}\left(
\re^{y(\tau)/2}- \re^{y(0)/2}\right).
\end{align}

\ukrainianpart
\title[Модель Мертона-Кармана.]%
{Континуальне представлення ядра оператора еволюції в моделі Мертона-Кармана}
\author[Л.Ф.~Блажиєвський, В.С.~Янішевський]{Л.Ф.~Блажиєвський\refaddr{label1},
        В.С.~Янішевський\refaddr{label2}}
\addresses{
 \addr{label1} Кафедра теоретичної фізики, Львівський Національний університет імені
Івана Франка, вул.~Драгоманова,~12,  79005~Львів, Україна
 \addr{label2} Кафедра економічної кібернетики та інноватики,
 Дрогобицький державний педагогічний університет імені Івана Франка, вул.~Лесі Українки,
  82100~Дрогобич, Україна}

\makeukrtitle

\begin{abstract}
\tolerance=3000%

В методі континуального інтегрування побудовано ядро оператора еволюції для
гамільтоніану Мертона-Кармана. На основі ядра отримана формула для ціни
опціону, що узагальнює відому формулу Блека-Шоулса. Вказано також на можливі
способи наближеного обчислення континуальних інтегралів.

\keywords континуальне інтегрування, ядро оператора еволюції, ціноутворення
опціонів,
формула Блека-Шоулса, модель Мертона-Кармана.%
\end{abstract}

\end{document}